\documentclass[conference]{IEEEtran}
\usepackage[justification=centering]{caption}
\usepackage{graphicx}
\usepackage{algorithm}
\usepackage{comment}
\usepackage[noend]{algorithmic}
\ifCLASSINFOpdf
\else
\fi
\begin{document}
\title{Leading Undergraduate Students to Big Data Generation
}

\author{\IEEEauthorblockN{Jianjun Yang}
\IEEEauthorblockA{Department of Computer Science \\and Information Systems~~~~~~~~~~\\
University of North Georgia\\
Email: jianjun.yang@ung.edu}
\and
\IEEEauthorblockN{Ju Shen}
\IEEEauthorblockA{Department of Computer Science\\
University of Dayton\\
Email: jshen1@udayton.edu}

}


%Telephone: (800) 555--1212, Fax: (888) 555--1212}


\maketitle

\section{Introduction}

People are facing a flood of data today. Data are being collected at unprecedented scale in many areas, such as networking\cite{xukunjie1}\cite{xukunjie2}\cite{Yang3}, image processing\cite{shenju1}\cite{shenju2}, visualization\cite{shenju3}, scientific computation, data base\cite{Wang1}\cite{Wang2}, and algorithms. The huge data nowadays are called Big Data. Big data is an all-encompassing term for any collection of data sets so large and complex that it becomes difficult to process them using traditional data processing applications.(Wikipedia 2015). New technologies and new forms are driving the Big Data development, with the global Internet population growing by 6.5\% averagely in the past three years and now representing two billion people. Almost everyone heard the term "Big data" nowadays. Big Data is used in a wide variety of applications, such as traffic patterns, purchasing behaviors, online video, and real-time inventory management. Consequently, there is a high demand of job positions on Big Data. In Georgia and Ohio, for example, a critical need exists for a highly qualified information technology (IT) workforce regarding Big Data.  There are over 4,000 vacant IT jobs in Georgia and Ohio that employers cannot readily fill related to Big Data(Monster.com and CareerBuilder.com, 2014).
         Although almost everyone heard the term "Big Data", many people, even undergraduate students in Computer Science major have poor understanding of what Big Data is.  Big Data is critical for students' current study and future career; hence many schools are training Big Data to students. However, it is extremely difficult to teach it because: First, manipulating data sets often requires massively parallel software running on tens, hundreds, or even thousands of servers. Second, there is no specific Big Data course in most schools. Many instructors met a lot of challenges when they teach Big Data to students.  The challenges on teaching and learning Big Data include analysis, capture, search, sharing, storage, transfer, visualization, and privacy violations.
         In this article, the authors present a unique way which uses network simulator and tools of image processing to train students abilities to learn, analyze, manipulate, and apply Big Data. Thus they develop students' hands-on abilities on Big Data and their critical thinking abilities. The authors' work is not merely to introduce "Big Data". Rather, their projects incorporated students in concept learning, research design, data collection, data manipulation, analysis, and problem solving of networking and multimedia. The authors provided students with two areas of applications. The first one is on web/mobile. A simulator was provided and student learned how to simplify Big Data in networking to a single computer program. The second one was on image processing. The authors used novel image based rendering algorithm with user intervention to generate realistic 3D virtual world. The learning outcomes are significant.

\section{Design and Conduct Relevant Experiments}
%%Recent recent of computer hardware and physical layer  have significantly improved the characteristics of network devices and IEEE radios.
%Shen el at. developed smart grid and renewable energy power stations, which dramatically saved the power of
%mesh devices\cite{shenshen1}\cite{shenshen2}. X. Zhang et al. proposed grid model that significantly boost the reliability of hardware and devices. \cite{zhangxianjun1}\cite{zhangxianjun2}\cite{zhangxianjun3}. In addition, intelligence makes the radios
%cognitive\cite{wangjj1}\cite{wangjj2}\cite{wangjj3}\cite{lile1}\cite{lile2}\cite{lile3}.
%Moreover, Kunjie et al. significantly contributed to the performance modeling technique\cite{xukunjie1}\cite{xukunjie2}\cite{xukunjie3}.
%They all provided solid foundation for wireless mesh networks.

The  literature  highlights  the  importance  of hands-on  activities  in  the  teaching  of  technologies\cite{Curto}. Hence the authors trained students Big Data by Projects. In their teaching experiences, they assigned projects on networking and image processing for three phases from easy to difficult.

\subsection{Phase 1: Reorganization of Big Data and Simplification from Big Network Data to a Simple Simulator}

Big Data is critical in Computer Science not only because it is an emerging technology, but also it is fundamental for students' future career. Some Computer Science scholars have generally gravitated toward introducing easy content under the assumption that the students would be more receptive to it. It is not true. If the goal of teaching Big Data is just to introduce the basic concepts, it would be an easy task by simplifying the course. However this could make students, especially those in computer science major, get bored easily with those trivial and superficial contents. Moreover, this teaching strategy prevents students from grasping the fundamentals concretely. However, it is difficult for students to learn abstract concepts of Big Data by merely taking classes from the instructor. To instill the joy of Big Data to students, the authors demonstrate interesting cases to stimulate students' learning interest.

         Appropriate teaching tools can effectively illustrate the theories of Big Data, which are abstract and often complicated to understand. When the authors taught Computer Network course, they studied the characteristics of wireless devices including laptops, iPads, iPhones and Android Phones. In order to consistently create an enthusiastic learning environment and facilitate student's success, they used simulator as a tool to introduce and simplify Big Data in networking. Simulations are an act of imitating the behavior of a physical or abstract system, such as an event, a situation, or a process that does or could exist\cite{Damassa}. Some scholars\cite{Maran} consider simulations as a perfect educational technique that creates learning by reproducing all or part of an event or situation. Theoretically, simulations could be created for any number of topics, courses, or programs in education. Some of   more  popular  simulations  are  offered  in  various academic  programs  including  business,  health  care,  and transportation.  Technology advances allow individuals to design self-placed simulations in their classrooms with limitless options. This has led to a full-fledged market for simulations in a  wide range of  areas  like  stock markets, roller coasters, and trucking. In this paper, the authors design a software program as a simulator for mobile networking.
         
         The author simplified big network data by simulators. The authors showed the real network topology and the presented the graphical interface of the designed simulator when teaching Big Data. When the authors introduced Big Data, they presented the scenario of connected network devices. Since many modern and popular devices are used in the scenario, it makes the class compelling and retain students' attentions.  High volume data are demonstrated from different aspects, such as their structures, transmissions, and representations among the network devices including its structure, transmission, and representation under the network devices. Then they introduced how to retrieve the critical content from the Big Data, such as IP addresses, locations, the resource capabilities \cite{Yang2}\cite{Yang4}\cite{Yang1}.  Afterwards, the teachers instructed students to practice manipulating Big Data through hands-on projects.  Students are guided to allocate to allocate the resources to the mobile devices by solving linear equations. They pinpointed areas where students can add nodes based on the properties of the heterogeneous devices, in order to increase the number of equations. This is the way to simplify the big network data. Students later implement the equations by programming and the results are displayed in the simulator. This provides undergraduate students a unique opportunity to use experimental technologies to be adaptively involved in learning complicated Big Data problems and understanding the abstract concepts.

\subsection{Phase 2: Development of Android App}

Marc  Prensky  (2001)\cite{Prensky}  created  a  powerful  summary when  he  said  games  offer  fun,  play,  rules,  goals, interactivity,  outcomes,  feedback,  conflict,  opposition,      problem solving, structure, flow, motivation, and pleasure. With such a list of benefits, it is a good idea to use smart phone for teaching in the classroom.

         The authors taught Android development for Big Data. App Inventor for Android is a new visual programming platform for creating mobile applications (apps) for smart phones. It was developed at Google Labs by a team led by MIT.  To developed apps in App Inventor students do not write code. Instead, they designed tools to visualize the app by using block based GUI for students to directly control the app's behaviors through interlocking components..  App Inventor aims to develop intuitive tools that facilitate novices to program in an enjoyable manner. App Inventor lets students create apps for smart phones. Given the popularity and ubiquity of mobile phones among today's generation of students, App Inventor seems to hold a great potential for attracting a new generation of students to problem-solving thinking to handle Big Data.
         
         Students found App Inventor very accessible and they learned how to develop apps of their own design quickly. Though the App looks simple, it actually incorporates a large amount of data with different formats (e.g. images, sounds, labels, etc.), and involves considerable control logics.  . Hence App is able to let students focus on problem solving on handling the big data rather than coding syntax.  The authors asked students to design some very interesting App projects. For example, they assigned the students to develop an interactive map of the attractions in Paris. When an attraction is clicked, its corresponding information will be displayed.  
         
         App is a good tool to develop students' problem-solving ability since it is not only easy to follow and reproduce already written apps, but also straight forward to develop completely new apps based on the principles acquired through the tutorials and demonstrations. Students progressed quickly from writing "Hello Kitty" to developing apps using database, interactive maps, client server communication, and other advanced concepts. Thus they know how to manipulate Big Data, even when they encounter problems.  Students were able to apply their programming skills to new types of problems including databases, client-server communication, images processing and algorithms.

\subsection{Phase 3:  Big Visual Data Editing System for Image Retrieval and Reconstruction}

This is the third phase to train students Big Data. The authors proposed an interactive system to operate on big visual data that supports online picture sharing or virtual 3D world navigation when they taught Interactive Media. Students got involved of the whole process of system development, such as coding, online image editing, and 3D model designing.

         With the explosive growth of internet and web-based cameras, billions of photographs are uploaded to the internet every day. The massive collections of imagery have inspired a wave of different applications on such large visual data. Part of the excitement in these areas is due to the facts that images are easy to take nowadays everywhere from our daily devices, like cell phones, tablets, and the efficient online access via WiFi or any phone network. Imagine building a virtual 3D world by taking the advantage of these large online images, such as the Google street view databases or the Flickr image collection. This system can provide virtual environment and immersive experience that allows users to walk freely in a re-constructed virtual world and view the scene from any arbitrary perspectives. In addition to its virtual reality value, as a photo warehouse, such system can also support large visual information. For example, for a travelling resort, people often take many pictures during the trips. However, sometimes the taken pictures may be less than satisfactory, such as the background scene is not fully captured or occluded by some objects. Some photo editing tools are available to improve the images. However, it could be a pain to modify the picture directly without any extra information, which often introduces noticeable artifacts. Things can become much easier, if there are additional available pictures taken from the same location at similar time. In such a way, travelers can share their experiences and enrich their photo collections from the large visual data.
         
         The authors assigned students with a series of projects which are on image retrieval, localization and reconstructing 3D geometry from a large, unordered collection of online images on landmarks and cities\cite{Irschara}\cite{Li}\cite{Hays}.  Because students have experience from the first two phases, the authors asked students to use image feature descriptors, such as SIFT or SURF, as the cue to identify similar images for clustering. Then based on the detected feature correspondences across multiple views, the scene geometry can be approximated estimated. The use of real photos not only supports realistic image synthesis with little user intervention, but raises the important issue of controlling and altering the representations. The students were really interested in the projects and happy to present their work to the instructors. Many results have demonstrated that, through training, students have developed the ability to use tools to render realistic view of novel images efficiently and accurately.

         The projects of this phase present an integrated research and educational program with two goals. The first goal of the phase is to produce new technologies on intuitive and interactive pictorial editing tools that allow undergraduates to manipulate and alter large visual data directly in high dimensions or temporal domain. The second goal of the phase is to expose the cutting edge technologies in Big Data processing, especially for visual data clustering and reconstruction to undergraduates, which can stimulate student interests in the related fields and promote their pursuit of careers. This phase is not only undergraduate oriented as many available software tools can be used straight away, such as the image matching APIs, 3D transformation tools, but also requires students to explore the core techniques and develop novel solutions on efficiently manipulating large visual data. During the phase, students had the chance to learn those well-established algorithms and state-of-the art Big Data technologies in image matching, 3D graphics, and data visualization. In some applications, people only need to know the outline of a car. Figure 5 shows the process to reduce the Big Data to represent a car to much smaller data that represents the outline of the car.

\section{Conclusion}
\label{conclusion}

Big Data is very important yet very difficult to teach for students. In this paper, the authors proposed an effective way to teach Big Data to students. They did not merely mechanically introduce the concepts of Big Data. Instead, they used concrete examples to illustrates Big Data to students gradually through three phases. They assigned students relevant projects to train their skills on handling Big Data, develop students' critical thinking and finally lead students to Big Data generation.



%\bibliographystyle{IEEEtran}

\end{document}